# Generalized structure of hadron-quark vertex function in Bethe-Salpeter framework: Applications to leptonic decays of V-mesons

Shashank Bhatnagar[1]
Department of Physics, Addis Ababa University, P.O.Box 101739, Addis Ababa, Ethiopia
and
Abdus Salam International Center for Theoretical Physics (ICTP), Trieste, Italy

Shi-Yuan Li [2]
Department of Physics, Shandong University, Jinan, 250100, P.R. China
and
Institute of Particle Physics, Huazhong Normal University, Wuhan, 430079, P. R. China

**Abstract:** We employ the framework of Bethe-Salpeter equation under Covariant Instantaneous Ansatz to study the leptonic decays of vector mesons. The structure of hadron-quark vertex function $\Gamma$ is generalized to include various Dirac covariants (other than $i\gamma.\varepsilon$) from their complete set. They are incorporated in accordance with a naïve power counting rule order by order in powers of the inverse of the meson mass. The decay constants for $\rho, \omega$ and $\phi$ mesons are calculated with the incorporation of leading order covariants.

1. Regular Associate, ICTP, shashank_bhatnagar@yahoo.com
2. lishy@sdu.edu.cn

## 1. Introduction:

Meson decays provide an important tool for exploring the structure of these simplest bound states in QCD and for studying the non-perturbative (long distance) behavior of strong interactions. Since the task of calculating hadron structure from QCD alone is very difficult, one relies on specific models of hadron dynamics to gain some understanding of hadronic structures at low energies. This study can most effectively be accomplished by applying a particular framework of hadron dynamics to a diverse range of phenomena. A number of studies in nonperturbative QCD have been carried out on pseudo-scalar mesons. However, the situation is different in case of vector mesons. Flavorless vector mesons play an important role in hadron physics due to their direct coupling to photons and thus provide an invaluable insight into phenomenology of electromagnetic couplings to hadrons. Thus a realistic description of vector mesons at the quark level of compositeness would be an important element in our understanding of hadron dynamics and reaction processes. There have been a number of recent studies [1-6] on processes involving strong decays, radiative decays and leptonic decays of vector mesons. Such studies offer a direct probe of hadron structure and help in revealing some aspects of the underlying quark-gluon dynamics that are complementary to what is learnt from pseudo-scalar mesons.

Thus in this paper we study leptonic decays of vector mesons (Hereafter, we use the short form V-mesons) which proceed via one photon annihilation of $q\bar{q}$ pair constituting V-meson through quark loop diagram (Fig.1). We employ a QCD oriented framework of Bethe-Salpeter Equation (BSE) under Covariant Instantaneous Ansatz (CIA) [7]. CIA is a Lorentz-invariant generalization of Instantaneous Approximation (IA). For $q\bar{q}$ system, CIA formulation [7] ensures an exact interconnection between 3D and 4D forms of the BSE. The 3D form of BSE serves for making contact with the mass spectra of hadrons, whereas the 4D form provides the $Hq\bar{q}$ vertex function $\Gamma(\hat{q})$ for evaluation of transition amplitudes. A BS framework under IA was also earlier suggested by Bonn group [8] where they have applied their formalism to various physical processes.

We had earlier employed the framework of BSE under CIA for calculation of decay constants [7, 9] of heavy-light pseudoscalar mesons and calculation of $f_\pi$ in $\pi^0 \to 2\gamma$ process. In a recent work we had evaluated the leptonic decays of V-mesons (such as $\rho, \omega, \phi$) [10] in this framework. However, one of the simplified assumptions in all these calculations was the fact that the BS wave function was restricted to have a single Dirac structure (e.g., $\gamma_5$ for P-mesons, $\gamma.\varepsilon$ for V-mesons etc.). However recent studies [2, 11-13] have revealed that various mesons have many different covariant structures in their wave functions whose inclusion was also found necessary to obtain quantitatively accurate observables [2].

Hence the present work grew out as a first attempt to develop and use a BS wave function for a hadron, which would incorporate more than a single Dirac structure for calculation of transition amplitudes for various processes. We first postulate and discuss a power counting rule for choosing various Dirac covariants from their complete set (see, e.g., [2, 11-13]) for V-mesons' wave functions in Section 2. In section 3 we calculate leptonic decay constants of V-mesons using the wave function developed in section 2. We conclude with discussions in Section 4.

## 2. Structure of generalized vertex function $\Gamma(\hat{q})$ in BSE under CIA:

We first outline the CIA framework taking the case of scalar "quarks" for simplicity. For a $q\bar{q}$ system with an effective kernel $K$ and 4D wave function $\Phi(P,q)$, the 4D BSE takes the form,

$$i(2\pi)^4 \Delta_1 \Delta_2 \Phi(P,q) = \int d^4q' K(q,q')\Phi(P,q'), \qquad (2.1)$$

where $\Delta_{1,2}$ are the inverse propagators of two scalar quarks. Hereafter, we use the free particle propagator, i.e.,

$$\Delta_{1,2} = m_{1,2}^2 + p_{1,2}^2, \qquad (2.2)$$

where $m_{1,2}$ are (effective) constituent masses of quarks. The 4-momenta of the quark and anti-quark, $p_{1,2}$, are related to the internal 4-momentum $q_\mu$ of the hadron by

$$p_{1,2\mu} = \hat{m}_{1,2} P_\mu \pm q_\mu. \qquad (2.3)$$

Here $P = p_1 + p_2$ is the total 4-momentum of the hadron with mass $M$. $\hat{m}_{1,2} = [1 \pm (m_1^2 - m_2^2)/M^2]/2$ are the Wightman-Garding (WG) definitions [9] of masses of individual quarks. We now use an ansatz on the BS kernel $K$ in Eq. (2.1), which is assumed to depend on 3D variables $\hat{q}_\mu, \hat{q}_\mu'$, i.e.,

$$K(q,q') = K(\hat{q},\hat{q}') \qquad (2.4)$$

where

$$\hat{q}_\mu = q_\mu - \frac{q.P}{P^2} P_\mu \qquad (2.5)$$

is observed to be orthogonal to the total 4-momentum $P$ (i.e., $\hat{q}.P = 0$) irrespective of whether the individual quarks are on-shell or off-shell. A similar form of the BS kernel was also suggested in ref. [8]. Hence the longitudinal component of $q_\mu$,

$$M\sigma = M \frac{q.P}{P^2} \qquad (2.6)$$

does not appear in the form $K(\hat{q},\hat{q}')$ of the kernel. For reducing Eq. (2.1) to the 3D form, we define a 3D wave function $\phi(\hat{q})$ as

$$\phi(\hat{q}) = \int_{-\infty}^{+\infty} Md\sigma \Phi(P,q). \qquad (2.7)$$

Substituting Eq.(2.7) in Eq.(2.1) with the definition of kernel in Eq.(2.4), we get a covariant version of the Salpeter Equation [14],

$$(2\pi)^3 D(\hat{q})\phi(\hat{q}) = \int d^3\hat{q}' K(\hat{q},\hat{q}')\phi(\hat{q}') ,\qquad (2.8)$$

where $D(\hat{q})$ is a 3D denominator function defined by

$$\frac{1}{D(\hat{q})} = \frac{1}{2\pi i}\int_{-\infty}^{+\infty}\frac{Md\sigma}{\Delta_1\Delta_2} \qquad (2.9)$$

whose value can be easily worked out by contour integration by noting the positions of the poles in the complex $\sigma$–plane (as shown in detail in Ref.[10]). We note that the RHS of Eq. (2.8) is identical to RHS of Eq. (2.1) by virtue of Eqs.(2.4) and (2.7). We thus have an exact interconnection between 3D wave function $\phi(\hat{q})$ and the 4D wave function $\Phi(P,q)$:

$$\Delta_1\Delta_2\Phi(P,q) = \frac{D(\hat{q})\phi(\hat{q})}{2\pi i} \equiv \Gamma(\hat{q}) \qquad (2.10)$$

where $\Gamma(\hat{q})$ is the BS vertex function under CIA. The exact interconnection between 3D and 4D forms of BSE under CIA is thus brought out. The 3D form serves for making contact with the mass spectrum of hadrons, whereas the 4D form provides the $Hq\bar{q}$ vertex function $\Gamma(\hat{q})$ which satisfies a 4D BSE with a natural off-shell extension over the entire 4D space (due to the positive definiteness of the quantity $\hat{q}^2 = q^2 - \frac{(q.P)^2}{P^2}$ throughout the entire 4D space) and thus provides a fully Lorentz-invariant basis for evaluation of various transition amplitudes through various quark loop diagrams. Due to these properties, this framework can be profitably employed not only for mass spectral predictions but also for evaluation of various transition amplitudes [7,9,10] all the way from low energies to high energies.

We now outline the framework of BSE under CIA for the case of fermionic quarks constituting a particular meson. The scalar propagators $\Delta_i^{-1}$ in the above equation are replaced by the proper fermionic propagators $S_F$. Now comes the problem of incorporation of relevant Dirac structures in vertex function $\Gamma(\hat{q})$. In this connection we wish to state that in applications of BSE under CIA until now, the $Hq\bar{q}$ vertex function was restricted to have a single Dirac structure. However recent studies [2,11-13] have revealed that various mesons have many different covariant structures in their wave functions and their inclusion was also found necessary to obtain quantitatively accurate observables [2]. Hence the present study attempts to incorporate other Dirac covariants in the structure of $Hq\bar{q}$ vertex function systematically. Thus to incorporate the relevant Dirac structures in the vertex function, we take guidance from some of the recent works [2, 11-13] where the transverse $Hq\bar{q}$ vertex function for a V-meson has been expressed as a linear combination of eight Dirac covariants [2, 11-13], $\Gamma_i^V$ (i=0,…,7), each multiplying a Lorentz scalar amplitude $F_i(q^2, q.P, P^2)$. Similarly the $Hq\bar{q}$ vertex functions for both pseudo-scalar and scalar mesons are expressible as a linear combinations of four Dirac covariants $\Gamma_i^P$ ($i=0,...,3$) and $\Gamma_i^S$ ($i=0,...,3$) respectively [2, 11-13]. (The discussion on pseudoscalar and scalar mesons will be relegated to a separate paper.) However, the choice of these covariants is not unique, as can also be seen from the choice of the eight covariants employed in Ref. [2, 11-13] for a V-meson. What we want to do is to find a "criterion" so as to systematically choose among these eight Dirac covariants. In this connection we wish to state that in Ref. [2, 11] for a V-meson, it was noticed that the covariant $\Gamma_1 = \hat{\gamma}_\mu$ (where $\hat{\gamma}_\mu = \gamma_\mu - P_\mu(P.\gamma)/P^2$ is the transverse projection of the four vector $\gamma_\mu$) was considered to be the most important one. Such 'leading order' covariants ($i\gamma.\varepsilon$ for V-meson, $\gamma_5$ for P-meson etc.) were used in our work so far for calculation of some transition amplitudes [7, 9, 10] (since it can be noticed that $\gamma.\varepsilon$ is the same as $\hat{\gamma}.\varepsilon$ on account of $P.\varepsilon = 0$ with $\varepsilon_\mu$ the polarization vector). These were also used in earlier calculations of various transition amplitudes and mass spectra of hadrons in BSE under Null-Plane Ansatz (NPA) [15]. On the other hand in Ref. [2, 11] only five of the eight covariants (ie. $\Gamma_1,...,\Gamma_5$) were considered to be important for calculation of vector meson masses and decay constants. Further in Ref. [2, 11], calculations have also been made for masses and decay constants of V-mesons for various other subsets of eight covariants (see table III of Ref.[2]).

On lines of these studies, we try to generalize the structure of $Hq\bar{q}$ vertex function for a V-meson

$$\Gamma(\hat{q}) = \frac{1}{2\pi i} N_V (i\gamma.\varepsilon) D(\hat{q}) \phi(\hat{q}) \tag{2.11}$$

which we had used earlier [10]. For incorporating the Dirac structures in the expression for $\Gamma(\hat{q})$ we take their forms as in Ref. [12]. We notice that in the expression for the CIA vertex function, $\Gamma(\hat{q})$ in Eq. (2.11), the factor $D(\hat{q})\phi(\hat{q})$ multiplying $i\gamma.\varepsilon$ is nothing but the Lorentz invariant momentum dependent scalar which depends upon $q^2, P^2$ and $q.P$ (see Sec. 3 of this paper) and thus has a certain dimensionality of mass. However, the Lorentz scalar amplitudes $\Gamma_i(q,P)$, ($i=0,1,2,...$) multiplying the various Dirac structures in Ref. [12] have different dimensionalities of mass. For adapting this decomposition to write the structure of vertex function $\Gamma(\hat{q})$, we re-express the $Hq\bar{q}$ vertex functions for V-meson (see Eq.(10) of Ref. [12]) by making these coefficients $\Gamma_i(q,P)$ dimensionless, weighing each covariant with an appropriate power of $M$, the meson mass. Thus each term in the expansion of $\Gamma(\hat{q})$ is associated with a certain power of $M$. In detail, we can express $\Gamma_\mu^V \varepsilon_\mu$ as

$$\Gamma_\mu^V \varepsilon_\mu = \Omega_\mu \varepsilon_\mu . \frac{1}{2\pi i} N_V D(\hat{q}) \phi(\hat{q}) ; \tag{2.12}$$

$$\Omega_\mu \varepsilon_\mu = i(\gamma.\varepsilon)A_0 + (\gamma.\varepsilon)(\gamma.P)\frac{A_1}{M} + [q.\varepsilon - (\gamma.\varepsilon)(\gamma.q)]\frac{A_2}{M}$$

$$+ i\frac{A_3}{M^2}[(\gamma.\varepsilon)(\gamma.P)(\gamma.q) - (\gamma.\varepsilon)(\gamma.q)(\gamma.P) + 2i(q.\varepsilon)(\gamma.P)]$$

$$+ (q.\varepsilon)\frac{A_4}{M} + i(q.\varepsilon)(\gamma.P)\frac{A_5}{M^2} - i(q.\varepsilon)(\gamma.q)\frac{A_6}{M^2} + (q.\varepsilon)[(\gamma.P)(\gamma.q) - (\gamma.q)(\gamma.P)]\frac{A_7}{M^3}$$

where $A_i$ ($i = 0,...,7$) are eight dimensionless and constant coefficients (why they can be constant will be explained in the following) to be determined. Now since we use constituent quark masses where the quark mass $m$ is approximately half of the hadron mass $M$, we can use the ansatz

$$q \ll P \sim M \qquad (2.13)$$

in the rest frame of the hadron. Then each of the eight terms in Eq.(2.12) receives suppression by different powers of $1/M$. Thus we can arrange these terms as an expansion in powers of $O(\frac{1}{M})$. We can then see in the expansion of $\Omega_\mu \varepsilon_\mu$ that the structures associated with the coefficients $A_0, A_1$ have magnitudes $O(\frac{1}{M^0})$ and are of leading order. Those with $A_2, A_3, A_4, A_5$ are $O(\frac{1}{M^1})$, while those with $A_6, A_7$ are $O(\frac{1}{M^2})$. This naïve power counting rule suggests that the maximum contribution to the calculation of any vector meson observable should come from the Dirac structures $i\gamma.\varepsilon$ and $(\gamma.\varepsilon)(\gamma.P)\frac{1}{M}$ associated with the constant coefficients $A_0$ and $A_1$ respectively. As a first attempt we take the form of the $Hq\bar{q}$ vertex function incorporating these leading order terms in expansion (2.12) and ignoring $O(\frac{1}{M^1})$ and $O(\frac{1}{M^2})$ terms for the moment and try to calculate the $V$-meson decay constants taking only these leading order terms. Thus we take the modified form of $Hq\bar{q}$ vertex function

$$\Gamma(\hat{q}) = [i\gamma.\varepsilon A_0 + (\gamma.\varepsilon)(\gamma.P)\frac{A_1}{M}].\frac{1}{2\pi i}N_V D(\hat{q})\phi(\hat{q}) \ . \qquad (2.14)$$

For the flavourless vector mesons, which are eigenstates of the charge parity, there is an extra restriction on the use of the Dirac structures [13]. In general, the coefficients $A_i$ of the Dirac structures could be functions of $q \cdot P$, hence can be written as a Taylor series in powers of $q \cdot P$. However, the coefficients here used are dimensionless. Hence the various terms in the series should be powers of $\frac{q \cdot P}{M^2}$, which is of order $O(\frac{1}{M})$. If we want to keep the leading contributions, as discussed above, we should only keep the zeroth-order terms of the Taylor series. This justifies the usage of $A_0$ and $A_1$ etc. as constant in the above equations. But now we choose only the C-even part of the coefficients, since only odd powers of q are C-odd. Hence only the proper C value Dirac structures can be used. The Dirac structures used in (2.14) are consistent with the charge parity of the flavourless vector mesons (the second term originates from $\sigma_{\mu\nu}P_\nu$). In this paper we will investigate the numerical results to this order.

In a similar manner one can express the full hadron-quark vertex function for a pseudoscalar, scalar and axial vector mesons, etc., in BSE under CIA, taking guidance from Ref. [12]. Then we can incorporate the Dirac structures according to the power counting rule. At the same time, the restriction by charge parity should also be respected.

From the above analysis of the structure of $Hq\bar{q}$ vertex function (in Eq.(2.14)) we notice that the structure of 3D wave function $\phi(\hat{q})$ as well as the form of the 3D BSE (Eq.(2.8)) are left untouched and have the same form as in our previous works which justifies the usage of the same form of the input kernel we used earlier. Now we briefly mention some features of the BS formulation employed. The structure of BSE is characterized by a single effective kernel arising out of a four-fermion lagrangian in the Nambu-Jonalasino [15-16] sense. The formalism is fully consistent with Nambu-Jona-Lasino [16] picture of chiral symmetry breaking but is additionally Lorentz-invariant because of the unique properties of the quantity $\hat{q}^2$, which is positive definite throughout the entire 4D space. The input kernel $K(q,q')$ in BSE is taken as one-gluon-exchange like as regards color $(\frac{1}{2}\vec{\lambda}_1 \cdot \frac{1}{2}\vec{\lambda}_2)$ and spin($\gamma_\mu^{(1)}\gamma_\mu^{(2)}$) dependence. The scalar function $V(q-q')$ is a sum of one-gluon exchange $V_{OGE}$ and a confining term $V_{Conf.}$ [9-10,15]. Thus

$$K(q,q') = \frac{1}{2}\vec{\lambda}^{(1)} \cdot \frac{1}{2}\vec{\lambda}^{(2)}V_\mu^{(1)}V_\mu^{(2)}V(q-q');$$

$$V_\mu^{(1,2)} = \pm 2m_{1,2}\gamma_\mu^{(1,2)};$$

$$V(\hat{q}-\hat{q}') = \frac{4\pi\alpha_S}{(\hat{q}-\hat{q}')^2} + \frac{3}{4}\omega_{q\bar{q}}^2 \int d^3\vec{r}[r^2(1+4a_0\hat{m}_1\hat{m}_2M^2r^2)^{-\frac{1}{2}} - \frac{C_0}{\omega_0^2}]e^{i(\hat{q}-\hat{q}').\vec{r}}; \qquad (2.15)$$

$a_0 = .028, C_0 = .29$.

The values of parameters $a_0, C_0$ have been calibrated to fit the meson mass spectra obtained by solving the 3D BSE [15]. Here in the expression for $V(\hat{q}-\hat{q}')$, the constant term $C_0/\omega_0^2$ is designed to take account of the correct zero point energies, while $a_0$ term ($a_0 \ll 1$) simulates an effect of an almost linear confinement for heavy quark sectors (large $m_1, m_2$), while retaining the harmonic form for light quark sectors (small $m_1, m_2$) [15]. This representation is thus asymptotically consistent with linear confinement (as is believed to be true for QCD) though the intervening length scales in light quark sectors give it an effectively harmonic appearance. Now comes to the problem of the 3D BS wave function. The ground state wave function $\phi(\hat{q})$ satisfies the 3D BSE, Eq. (2.8) on the surface P.q = 0, which is appropriate for making contact with O(3)-like mass spectrum [15]. Its fuller structure (described in Ref.[15]) is reducible to that of a 3D harmonic oscillator with coefficients dependent on the hadron mass M and the total quantum number N. The ground state wave function $\phi(\hat{q})$ deducible from this equation thus has a gaussian structure [7,9,10] and is expressible as:

$$\phi(\hat{q}) \approx e^{-\hat{q}^2/2\beta^2} \quad . \tag{2.16}$$

In the structure of $\phi(\hat{q})$ in Eq. (2.16), the parameter $\beta$ is the inverse range parameter which incorporates the content of BS dynamics and is dependent on the input kernel $K(q,q')$. The structure of $\beta$ is given in Section 3.

The ansatz employed for the spring constant $\omega_{q\bar{q}}^2$ in Eq. (2.15) is [9-10,15]:

$$\omega_{q\bar{q}}^2 = 4\hat{m}_1\hat{m}_2 M \omega_0^2 \alpha_S(M^2) \tag{2.17}$$

where $\hat{m}_1, \hat{m}_2$ are the Wightman-Garding definitions of masses of constituent quarks defined earlier.

This approach is analogous to other studies [1-3, 11], which use generalized ladder approximation for studying bound state problems in QCD. We now give the calculation of leptonic decays of vector mesons in the framework discussed in this section.

### 3. Leptonic Decays of Vector Mesons:

Vector meson decay proceeds through the loop diagram shown in Fig. 1. The coupling of a vector meson to the photon is expressed via a dimensionless coupling constant $g_V$ which can be described as

$$\frac{M^2}{g_V}\varepsilon_\mu(P) = <0|\overline{Q}\hat{\Theta}\gamma_\mu Q|V(P)> \tag{3.1}$$

(where $Q$ is the flavor multiplet of quark field and $\hat{\Theta}$ is the quark electromagnetic charge operator), which can in turn be expressed as a loop integral [15],

$$\frac{M^2}{g_V}\varepsilon_\mu = \sqrt{3}e_Q \int d^4q Tr[\Psi_V(P,q)i\gamma_\mu] \tag{3.2}$$

where $e_Q^2 = \frac{1}{2}, \frac{1}{9}, \frac{1}{18}$ for $V = \rho, \phi, \omega$ respectively, and the polarization vector of V-meson $\varepsilon_\mu$ satisfying $\varepsilon.P = 0$.

FIG 1.

Bethe-Salpeter wave function $\Psi(P,q)$ for a V-meson is expressed as

$$\Psi(P,q) = S_F(p_1)\Gamma(\hat{q})S_F(-p_2);$$

$$S_F(p_1) = -i\frac{(m_1 - i\gamma.p_1)}{\Delta_1}; S_F(-p_2) = -i\frac{(m_2 + i\gamma.p_2)}{\Delta_2}. \quad (3.3)$$

In the following calculation, we only take the leading order terms in the structure of hadron-quark vertex function $\Gamma(\hat{q})$ as in Eq. (2.14). $S_F$ are the fermionic propagators for the two constituent quarks of the hadron and the nonperturbative aspects enter through the $\Gamma(\hat{q})$. Using $\Psi(P,q)$ from Eq.(3.3) and the structure of $Hq\bar{q}$ vertex function $\Gamma(\hat{q})$ from Eq.(2.14), evaluating the trace over $\gamma$-matrices and noting that only the components of $p_{1\mu}$ and $p_{2\mu}$ in the direction of $\varepsilon_\mu$ will contribute to $g_V$, we further make the replacements $p_{2\mu} \to (p_2.\varepsilon)\varepsilon_\mu$, $p_{1\mu} \to (p_1.\varepsilon)\varepsilon_\mu$. We can then write Eq. (3.2) as

$$\frac{M^2}{g_V} = \sqrt{3}e_Q \int d^3\hat{q}\frac{N_V D(\hat{q})\phi(\hat{q})}{2\pi i}I;$$

$$I = \int \frac{Md\sigma}{\Delta_1\Delta_2}[-4(m_1m_2 - \frac{1}{3}p_1.p_2)A_0 + (4m_2 P.p_1 + 4m_1 P.p_2)\frac{A_1}{M}] \quad (3.4)$$

where $p_1.p_2$ can be expressed as

$$p_1.p_2 = -\hat{m}_1\hat{m}_2 M^2 - \hat{m}_2\sigma M^2 - \hat{q}^2 + M^2\hat{m}_1\sigma + M^2\sigma^2 = m^2 - \frac{1}{2}(\Delta_1 + \Delta_2 + M^2) \quad (3.5)$$

taking $\hat{m}_1 = \hat{m}_2 = \frac{1}{2}$. With

$$p_1.P = \frac{1}{2}(\Delta_1 - \Delta_2 - M^2),$$

$$p_2.P = \frac{1}{2}(-\Delta_1 + \Delta_2 - M^2), \quad (3.6)$$

the integral $I$ in Eq. (3.4) can be expressed as

$$I = \int \frac{Md\sigma}{\Delta_1\Delta_2}\{[\frac{1}{6}(\Delta_1 + \Delta_2) + (\frac{1}{6}M^2 + \frac{2}{3}m^2)]A_0 + 4mMA_1\}. \quad (3.7)$$

Carrying out integration over $d\sigma$ by method of contour integration in the complex $\sigma$-plane (for details see Ref.[10]) and noting that V-meson decay constant $f_V = M/(e_Q g_V)$ [1], we can write

$$f_V = \frac{4\sqrt{3}}{M}N_V\left[\int d^3\hat{q}\phi(\hat{q})\{[(\frac{M^2}{6} + \frac{2m^2}{3}) + \frac{1}{6}D_0(\hat{q})]A_0 + 4mMA_1\}\right] \quad (3.8)$$

where the relationship between the functions $D_0(\hat{q})$ and $D(\hat{q})$ (see Ref.[10]) is

$$D(\hat{q}) = \frac{D_0(\hat{q})}{(\frac{1}{2\omega_1} + \frac{1}{2\omega_2})}; \qquad D_0(\hat{q}) = (\omega_1 + \omega_2)^2 - M^2 \qquad ; \qquad \omega_{1,2}^2 = m_{1,2}^2 + \hat{q}^2$$

(3.9)

In Eq. (3.8), $4mMA_1$ is the contribution to the integrand for $f_V$ due to the presence of the additional Dirac structure $\frac{(\gamma.\varepsilon)(\gamma.P)}{M}$ in the vertex function $\Gamma(\hat{q})$. This additional term appears with the same sign as the contribution $[(\frac{M^2}{6} + \frac{2m^2}{3}) + \frac{1}{6}D_0(\hat{q})]$ from the Dirac structure $i\gamma.\varepsilon$. Thus this should have the effect of raising the values of decay constants obtained in earlier calculation [10].

The structure of the parameter $\beta$ in $\phi(\hat{q})$ appearing in Eq.(3.8) as well as other input parameters $\omega_0, \Lambda, C_0$ are borrowed from Ref. [9, 10,15] since it can be easily shown that the 3D form of BSE (responsible for spectroscopic predictions and which also controls the parametric structure of 3D wave function $\phi(\hat{q})$ and hence $\beta$ and also fixes the values of parameters $\omega_0, \Lambda, C_0$) under CIA has a structure which is formally equivalent to the 3D BSE under NPA (see Ref.[9,10,17]). Further the exact correspondence between the 3-vector $\vec{q}$ of NPA and the 4-vector $\hat{q}_\mu$ of CIA on the surface $q.P = 0$ formally preserves [17] the algebraic structure of 3D BSE and hence of the spectroscopic predictions of NPA (see also Ref.[10]). Thus quantitative details of BS model under CIA in respect of spectroscopy can be directly taken over from NPA formalism. The structure of parameter $\beta$ is thus taken from Ref. [9,10,15] as,

$$\beta^2 = (2\hat{m}_1\hat{m}_2 M \omega_{q\bar{q}}^2 / \gamma^2)^{1/2};$$

$$\gamma^2 = 1 + \frac{2\omega_{q\bar{q}}^2 C_0}{M\omega_0^2}; \qquad (3.10)$$

$$\alpha_S(Q^2) = \frac{12\pi}{33-2f}\left[\ln\frac{Q^2}{\Lambda^2}\right]^{-1}$$

where $\omega_{q\bar{q}}^2$ is expressed as in Eq. (2.17), and the input parameters of the model are taken as $\omega_0 = 0.158 GeV, C_0 = 0.29, \Lambda = 0.20 GeV$ [9, 10,15].

To calculate the normalization factor $N_V$, we use the current conservation condition [7],

$$2iP_\mu = (2\pi)^4 \int d^4q Tr[\overline{\Psi}(P,q)(\frac{\partial}{\partial P_\mu} S_F^{-1}(p_1))\Psi(P,q)S_F^{-1}(-p_2)] + (1 \leftrightarrow 2), \qquad (3.11)$$

where the momentum of constituent quarks can be expressed as

$$p_{1,2\mu} = P_\mu(\hat{m}_1 \pm \sigma) \pm \hat{q}_\mu. \qquad (3.12)$$

Taking the derivatives of inverse of propagators of constituent quarks with respect to the total 4-momentum $P_\mu$, taking $\Psi(P,q)$ from Eq. (3.3), evaluating trace over the $\gamma$-matrices and following usual steps, we can express Eq. (3.11) as

$$N_V^{-2} = i\frac{(2\pi)^2}{2}\int d^3\hat{q} D^2(\hat{q})\phi^2(\hat{q}).I_1 \ + \ (1 \leftrightarrow 2) \ ; \qquad (3.13)$$

$$I_1 = \int_{-\infty}^{+\infty} \frac{Md\sigma}{\Delta_1^2 \Delta_2}(\hat{m}_1 + \sigma)\{\frac{2}{3}iA_0^2[\frac{2}{M^2}\Delta_1^2 + \frac{1}{M^2}\Delta_2^2 + (M^2 + 4m_1^2) + \frac{2}{M^2}(2m_1^2 + M^2)\Delta_2 - \frac{3}{M^2}\Delta_1\Delta_2$$

$$+ \frac{1}{M^2}(-4m_1^2 + 3M^2)\Delta_1]$$

$$+ \frac{2}{3}iA_1^2[\frac{1}{M^2}\Delta_1^2 - \frac{4}{M^2}\Delta_2^2 + \frac{3}{M^2}\Delta_1\Delta_2 - \frac{8}{M^2}m_1^2\Delta_1 + \frac{1}{M^2}(8m_1^2 + M^2)\Delta_2 + 2(4m_1^2 + M^2)]\}$$

where all the cross terms involving $A_0 A_1$ cancel off. In the above expression terms like $\Delta_1^2, \Delta_2^2$ and $\Delta_1\Delta_2$ give divergent contributions due to the presence of the factor $(\hat{m}_1 + \sigma)$ in the integrand. To overcome this problem, one can consider, following Ref. [18], the charge to be concentrated on one of the quark lines (say $p_1$). This may amount to taking the derivative with respect to $p_1$ (instead of $P$) [18] in the expression for $N_V^{-2}$ in Eq.(3.11). Thus we can express $N_V^{-2}$ as,

$$N_V^{-2} = i\frac{(2\pi)^2}{2}\int d^3\hat{q} D^2(\hat{q})\phi^2(\hat{q}).I_1 \ + \ (1 \leftrightarrow 2) \ ; \qquad (3.14)$$

$$I_1 = \int_{-\infty}^{+\infty} \frac{Md\sigma}{\Delta_1^2 \Delta_2}\{\frac{2}{3}iA_0^2[\frac{2}{M^2}\Delta_1^2 + \frac{1}{M^2}\Delta_2^2 + (M^2 + 4m_1^2) + \frac{2}{M^2}(2m_1^2 + M^2)\Delta_2 - \frac{3}{M^2}\Delta_1\Delta_2$$

$$+ \frac{1}{M^2}(-4m_1^2 + 3M^2)\Delta_1]$$

$$+ \frac{2}{3}iA_1^2[\frac{1}{M^2}\Delta_1^2 - \frac{4}{M^2}\Delta_2^2 + \frac{3}{M^2}\Delta_1\Delta_2 - \frac{8}{M^2}m_1^2\Delta_1 + \frac{1}{M^2}(8m_1^2 + M^2)\Delta_2 + 2(4m_1^2 + M^2)]\}$$

where we have been able to get rid of the factor $(\hat{m}_1 + \sigma)$ and thus the integration can be carried out over all the terms in the above expression. The integration over $d\sigma$ over various terms in Eq. (3.14) which can be carried out (taking $m_1 = m_2$) by noting the pole positions in complex $\sigma$-plane gives us:

(i) $\int \frac{Md\sigma}{\Delta_1^2 \Delta_2}\Delta_1 = 2\pi i[\frac{1}{D(\hat{q})}]$;

(ii) $\int \frac{Md\sigma}{\Delta_1^2 \Delta_2}\Delta_2 = 2\pi i[\frac{2}{(2\omega)^3}]$;

(iii) $\int \frac{Md\sigma}{\Delta_1^2 \Delta_2}\Delta_1\Delta_2 = 2\pi i[\frac{1}{2\omega}]$;

(iv) $\int \frac{Md\sigma}{\Delta_1^2 \Delta_2} = 2\pi i[\frac{M^2 - 12\omega^2}{4\omega^3(M^2 - 4\omega^2)^2}]$;

(v) $\int \frac{Md\sigma}{\Delta_1^2 \Delta_2} \Delta_2^2 = 2\pi i[\frac{\omega^2 - \frac{1}{2}M^2}{\omega^3}]$. (3.15)

(vi) $\int \frac{Md\sigma}{\Delta_1^2 \Delta_2} \Delta_1^2 = 2\pi i[\frac{1}{2\omega}]$.

Hence we get

$$N_V^{-2} = i\frac{(2\pi)^2}{2}\int d^3\hat{q} D^2(\hat{q})\phi^2(\hat{q}).I_1 + (1 \leftrightarrow 2) ;$$

$$I_1 = \frac{2}{3}A_0^2\{\frac{2}{M^2}(\frac{1}{2\omega}) + \frac{1}{2M^2}(\frac{\omega^2 - M^2/2}{\omega^3}) + (M^2 + 4m^2)(\frac{M^2 - 12\omega^2}{4\omega^3(M^2 - 4\omega^2)^2})$$

$$+ 2(\frac{2}{M^2}m^2 + 1)(\frac{2}{(2\omega)^3} - \frac{3}{M^2}\frac{1}{2\omega}) + (3 - \frac{4}{M^2}m^2)(\frac{1}{D(\hat{q})})\}$$

$$+ \frac{2}{3}A_1^2\{\frac{1}{M^2}(\frac{1}{2\omega}) - \frac{2}{M^2}(\frac{\omega^2 - M^2/2}{\omega^3}) + \frac{3}{M^2}(\frac{1}{2\omega}) - 8\frac{m^2}{M^2}(\frac{1}{D(\hat{q})}) + (1 + 8\frac{m^2}{M^2})(\frac{2}{(2\omega)^3})$$

$$+ 2(4m^2 + M^2)(\frac{M^2 - 12\omega^2}{4\omega^3(M^2 - 4\omega^2)^2})\}$$

(3.16)

We have thus evaluated the general expressions for $f_V$ (Eq.(3.8)) and $N_V$ in framework of BSE under CIA, with Dirac structure $(\gamma.\varepsilon)(\gamma.P)/M$ introduced in the $Hq\bar{q}$ vertex function besides $i\gamma.\varepsilon$. We see that so far the results are independent of any model for $\phi(\hat{q})$. However, for calculating the numerical values of these decay constants one needs to know the constant coefficients $A_0$ and $A_1$ which are associated with the Dirac structures $i(\gamma.\varepsilon)$ and $(\gamma.\varepsilon)(\gamma.P)/M$ respectively. The relative value, $\frac{A_1}{A_0}$ is a free parameter without any further knowledge of the meson structure in the framework discussed above. In general, one can incorporate all the Dirac structures (i.e. associated with orders e.g. $O(1/M)$ etc. besides the leading orders $O(1/M^0)$) along with their respective coefficients $A_i$ in a Taylor's series of $\frac{q.P}{M^2}$ to express the BS wave function and calculate the decay constants $f_V$ of vector mesons to the desired order. However we wish to mention that even if we wish to study decay constants to order $O(1/M)$, we introduce four additional parameters $A_2,...,A_5$. To fit all these parameters ($A_0,...,A_5$) would lead to the necessity of taking data of more types of vector mesons which includes the heavy quark vector mesons (such as $J/\Psi$). To have a "Global analysis" and best fit for spectra of both light and heavy quark vector mesons along with their decay constants, we may face the problem of reparametrizing the input BS kernel. This is further supported by the fact that $f_V$ does not have a real solution for $\Psi$ with the same set of input BS model parameters when leading order coefficients $A_0, A_1$ are used. However as a first step we only vary this parameter $A_1/A_0$ at the lowest order to see the effects of introducing the Dirac structure $(\gamma.\varepsilon)(\gamma.P)/M$. Further, from the expression for $f_V$ (Eq.(3.8)), with constituent quark mass $m \sim \frac{1}{2}M$, we can notice that $\frac{1}{6}A_0$ and $2A_1$ are of about the same order if we consider that both the Dirac structures $i(\gamma.\varepsilon)$ and $(\gamma.\varepsilon)(\gamma.P)/M$ give the leading contribution. This suggests that as a rough estimate $\frac{A_1}{A_0} \approx .083$. We calculate $f_V$ for $\rho, \omega$ and $\phi$ mesons for values of $\frac{A_1}{A_0}$ in the range 0.06-0.10. This range however does not mean any preference but is adequate to show the dependence of $f_V$ on $A_1/A_0$. The results are given in Table I along with those of other models and experimental data [20]. It is seen that the numerical values of these decay constants in BSE under CIA improve when Dirac structure $(\gamma.\varepsilon)(\gamma.P)/M$ is introduced in the vertex function in comparison to the values calculated with only $i\gamma.\varepsilon$. Further discussions are in Sec. 4.

Table I

## 4. Discussions:

In this paper we have first postulated and discussed in detail a naïve power counting rule for incorporation of various Dirac covariants in the wave functions of different mesons. We then calculate $f_V$ for vector mesons with equal mass quarks ($\rho, \omega, \phi$) in the framework of Bethe-Salpeter Equation under Covariant Instantaneous Ansatz, using the hadron-quark vertex function $\Gamma(\hat{q})$ in Eq. (2.14). It is seen that the values of Decay constants can improve considerably when the Dirac structure $(\gamma.\varepsilon)(\gamma.P)/M$ is introduced in the vertex function, with tuned parameter $\frac{A_1}{A_0}$ and come close to the results of some recent calculations [2, 11, 19] as well as the experimental results [20].

In this connection we wish to state that in a recent work [2], the Leptonic decay constants have been calculated for light vector mesons $\rho, \omega, \phi$ and K* within a ladder-rainbow truncation of coupled Dyson-Schwinger and Bethe-Salpeter equations with a model 2-point gluon function using $Vq\bar{q}$ vertex function $\Gamma_V$ to be a linear combination of eight dimensionless orthogonal Dirac covariants [2, 11] $T_\mu^i(q,P)$. Each covariant multiplies a scalar amplitude $F_i(q^2, q.P, P^2)$ for three different parameter sets for effective interactions. However, it was observed that only **five** of the eight covariants (i.e., $T_\mu^1,...,T_\mu^5$) are important [2, 11] for the calculation of vector meson masses and decay constants. Of these, the first covariant $T_\mu^1 = \hat{\gamma}_\mu$ (where $\hat{\gamma}_\mu = \gamma_\mu - P_\mu(P.\gamma)/P^2$ is the transverse projection of the four vector $\gamma_\mu$) was found to be the most important one [2, 11]. We wish to state here that $T_\mu^1 = \hat{\gamma}_\mu$ is precisely one of the covariant used by us in this work since it can be noticed that $\gamma.\varepsilon$ (used by us in the structure of $\Gamma$ in Eq. (3.3)) is the same as $\hat{\gamma}.\varepsilon$ on account of $P.\varepsilon = 0$. The decay constants (for $\rho$ and $\phi$) calculated in this model [2, 11] for one of the parameter sets: $\omega = 0.4 GeV, D = 0.93 GeV^2$ using the first covariant, $T_\mu^1$ are $f_\rho = 200 MeV$ and $f_\phi = 220 MeV$. In our CIA model (using the same Dirac covariant) we obtain $f_\rho = 141.2 MeV$ which is on the lower side and $f_\phi = 191.3 MeV$ which is somewhat closer to the value obtained in Ref. [2, 11]. The value of $f_\rho$ calculated in Ref. [1] using Dyson-Schwinger equation model (which uses a single Dirac covariant in the structure of the $Vq\bar{q}$ vertex function as in the case of our earlier calculation under CIA formulation [10]) is $f_\rho = 163 MeV$ which is closer to the corresponding value of $f_\rho = 141.2 MeV$ calculated in BSE under CIA when only $i\gamma.\varepsilon$ is used.

However the values of these decay constants calculated using first five covariants $T_\mu^1,...,T_\mu^5$ in Ref. [2,11] comes out to be $f_\rho = 199 MeV$ and $f_\phi = 250 MeV$. While if all the eight covariants are used, the values of the corresponding decay constants come out to be $f_\rho = 207 MeV$ and $f_\phi = 259 MeV$ [2, 11]. These results suggest that the values of decay constants $f_V$ calculated in our model (using only the covariant $\gamma.\varepsilon$) are likely to improve if Dirac structure other than $\gamma.\varepsilon$ are included in the BS vertex function employed in our work. This is precisely the overall result we notice in our model in respect of evaluation of $f_V$ values for $\rho, \omega$ mesons when $\gamma.\varepsilon\gamma.P/M$ is inserted into the vertex function. These decay constant values (specially for $\rho$ and $\phi$ mesons) go up and come close to experimental results [20], where it is to be mentioned that for rough comparison with experiment we can take $f_V$ values for the ratio $A_1/A_0$ in the range 0.08 – 0.09 due to reasons given in Section 3.

However, the value of $f_\phi$ gets too large. It is seen that only when $A_1/A_0$ is reduced to 0.02, we reproduce the experimental value of $f_\phi = 228 MeV$. This may suggest that whereas the structures $(i\gamma.\varepsilon)$ and $(\gamma.\varepsilon)(\gamma.P)/M$, which happen to be the leading order terms in expansion (2.12) are sufficient for the calculation of decay constants of the lightest vector mesons, $\rho$ and $\omega$, they are not sufficient for the calculation of decay constants of heavier vector mesons like $\phi$, for which one would have to go to next to leading order terms (i.e., $O(1/M), O(1/M^2)$) in Eq. (2.12). The reason is that the relative momentum may not be small and the assumption $q << P \sim M$ may be worse for heavier of the light quark mesons. However, the power counting rule discussed in this paper is a systematic way to take account of all the possible Dirac structures in a complete set of bases. Only the fact is that to consider more Dirac structures, the data of more kinds of mesons have to be considered to fix the coefficients of the Dirac structures as parameters. As mentioned at the end of section 3, this may also lead us to consider heavy quark data for which one may have to reparametrize the input BS kernel.

We also want to mention that these Dirac covariants other than $\gamma_\mu$ might also be important for the study of processes involving large $q^2$ [13]. It would be interesting to study the effect of inclusion of other Dirac covariants in expression for $\Gamma(\hat{q})$ used in our BSE formalism as well as the effect it would have on the numerical predictions in various other calculations such as radiative decays ($V \to P\gamma$, $V \to PP\gamma$) and strong decays ($V \to PP$) of vector mesons.

Acknowledgements:
This work originated from discussions between both the authors at ICTP. It was done during the Associateship visit of one of the authors (SB) to ICTP in May-July 2005. We would like to thank ICTP for hospitality. SB also thanks Swedish International Development Agency (Sida) for financial support. SL is supported in part by National Natural Science Foundation of China (NSFC) with Grant No.10205009.

Note added:
After this study was completed, our attention was drawn to the fact that recently an instantaneous BS equation incorporating the exact propagators has been proposed [21]. However, to incorporate the full propagator in numerical calculation is not straightforward. Thus we leave the discussion of this topic in future work.

**Table I**: Calculated values (in $MeV$) of leptonic decay constants $f_V$ ($V = \rho, \omega, \phi$) in BSE under CIA for range of values of ratio $A_1/A_0$ from 0.06 to 0.10. The decay constants are calculated from data [20] from $\Gamma_{V \to e^+e^-} = \frac{4\pi}{3} \frac{\alpha_{em}^2 e_q^2}{M} f_V^2$. The masses of hadrons are also used from [20]. The values of constituent quark masses used are $m_{u,d} = 300 MeV$, $m_s = 540 MeV$. Comparisons are done with results obtained from other models and the experimental values.

|  |  | $f_\rho$ | $f_\omega$ | $f_\phi$ |
|---|---|---|---|---|
| BSE-CIA (using both the covariants $i\gamma.\varepsilon$ and $(\gamma.\varepsilon)(\gamma.P)/M$) for the ratio: $A_1/A_0$ | .06 | 188.3 | 183.8 | 283.9 |
|  | .07 | 197.4 | 192.7 | 295.2 |
|  | .08 | 206.4 | 201.5 | 306.4 |
|  | .09 | 215.3 | 210.3 | 317.1 |
|  | .10 | 224.4 | 219.5 | 328.2 |
| BSE- CIA (using only the covariant $i\gamma.\varepsilon$) |  | 142.2 | 137.0 | 191.3 |
| BSE-NPA[15] |  | 204 | 208 | 372 |
| SDE [2, 11] for parameter set $\omega = .4 GeV$, $D = .93 GeV^2$ and covariants: (i) $T_\mu^1 = \hat{\gamma}_\mu$ (ii) $T_\mu^1,...,T_\mu^5$ (iii) $T_\mu^1,...,T_\mu^8$ |  | 200 199 207 |  | 220 250 259 |
| QCD Sum Rule Model [19] |  |  |  | $230 \pm 25$ |
| SDE[1] |  | 163 |  |  |
| Expt. Results [20] |  | $220.9 \pm 1.7$ | $194.6 \pm 3.2$ | $228.45 \pm 2.87$ |

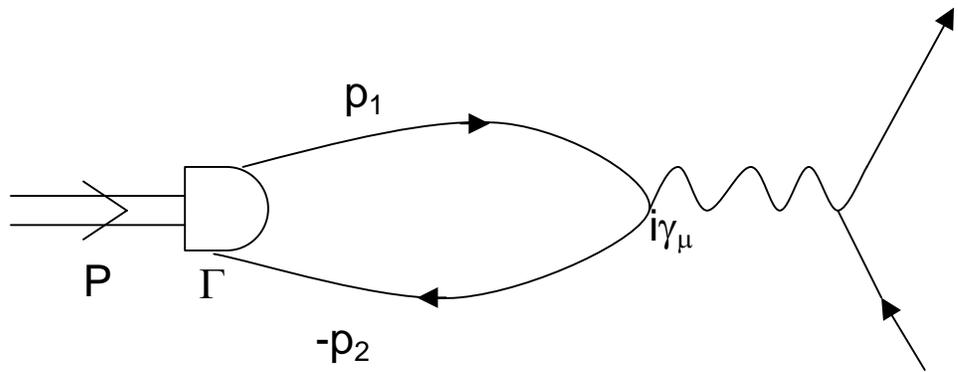

Fig.1: Quark loop diagram for leptonic decay of V-meson